\documentclass[aps,prb,twocolumn,groupedaddress,floats,showpacs]{revtex4}
\usepackage{latexsym}
\usepackage{dcolumn}
\usepackage[dvips]{graphicx}
\usepackage{amssymb}
\usepackage{graphics}
\usepackage{amsmath}
\usepackage{epsf}

\newcommand{\vk}{{\bf k}}
\newcommand{\vq}{{\bf q}}
\newcommand{\ve}{\varepsilon}

\begin{document}
\author{E. H. Hwang and S. Das Sarma}
\title{Acoustic phonon scattering limited carrier mobility in 2D
  extrinsic graphene}
\affiliation{Condensed Matter Theory Center, Department of Physics, 
University of Maryland, College Park, MD 20742-4111}
\date{\today}
\begin{abstract}
We theoretically calculate the phonon scattering limited electron
mobility in extrinsic (i.e. gated or doped with a tunable and finite
carrier density) 2D graphene layers as a function of temperature $(T)$ and
carrier density $(n)$. We find a temperature dependent phonon-limited
resistivity $\rho_{ph}(T)$ to be linear in temperature for $T\agt 50
K$ with the room temperature intrinsic mobility reaching values above
$10^5$ cm$^2/Vs$. We comment on the low-temperature
Bloch-Gr\"{u}neisen behavior where $\rho_{ph}(T) \sim T^4$ for unscreened
electron-phonon coupling.
\end{abstract}
\maketitle

\section{Introduction}

Low temperature carrier transport properties of 2D graphene layers have
been of great current interest to both experimentalists
\cite{Geim1,Zhang,Tan,Fuhrer,review} and theorists
\cite{Ando,Nomura,Falko,Hwang1,Adam} 
alike ever since the possibility of fabricating stable gated 2D
graphene monolayers on SiO$_2$ substrates and measuring the
density-dependent conductivity of the 2D chiral graphene carriers were
demonstrated \cite{Geim1}. Much of the early interest focused
understandably on the 
important issues of the scattering mechanisms limiting the
low-temperature conductivity and the associated graphene ``minimal
conductivity'' at the charge neutral (``Dirac'') point. 
One of the dominant low-temperature scattering 
mechanisms \cite{Tan,Fuhrer,Hwang1,Adam}  in graphene is that due
to screened Coulomb scattering by 
unintended charged impurities invariably present in the (mostly
SiO$_2$) substrate (and the substrate-graphene interface) although
short-range scattering by neutral defects also contributes,
particularly at high carrier densities in high-mobility samples. It
has therefore been argued \cite{Hwang1,Adam} that gated graphene
layers are similar to 2D 
electron systems in confined semiconductor structures (e.g. Si
inversion layers, GaAs heterostructures and quantum wells) where also
long-range charged impurity scattering dominates low-temperature ohmic
transport with short-range (e.g. interface roughness) scattering
playing a role at high carrier densities. \cite{AndoRMP,Manfra}

Given the exciting technological context of graphene as a prospective
electronic transistor material for future applications, the question
therefore naturally arises about the limiting value of the {\it
  intrinsic} room-temperature graphene mobility if all extrinsic
scattering mechanisms, e.g. charged impurities, neutral defects,
interface roughness, graphene ripples etc., can be eliminated from the
system. This question is more than of academic interest since serious
experimental efforts are underway \cite{Stomer} to eliminate charged
impurities from 
graphene by using different substrates or by working with free
standing graphene layers without any substrates. It is also noteworthy
that the systematic elimination of charged impurity scattering through
modulation doping and materials improvement in the MBE growth
technique has led to an astonishing 3,000-fold enhancement in the
low-temperature 2D GaAs electron mobility from $10^4$ cm$^2/Vs$ in 1978
to $30\times 10^6$ cm$^2/Vs$ in 2000, future enhancement to 100
million cm$^2/Vs$ mobility is anticipated \cite{Pfeiffer}
in the next few years.

One great advantage of graphene over high-mobility 2D GaAs systems is
that the lack of strong long-range polar optical phonon scattering,
which completely dominates \cite{Kawamura1} the room-temperature
GaAs mobility ($\sim 2,000$ 
cm$^2/Vs$), in graphene should lead to very high intrinsic room
temperature graphene mobility, limited only by the weak
deformation-potential scattering from the thermal lattice acoustic
phonons. In this work, we calculate the temperature-dependent 2D
graphene mobility limited only by the background lattice acoustic
phonon scattering. We find that room-temperature intrinsic (i.e. just
phonon-limited) graphene mobility surpassing $10^5$ cm$^2/Vs$ is
feasible using the generally accepted values in the literature for the
graphene sound velocity and deformation coupling. There is some
uncertainty in the precise value of the electron-phonon deformation
potential coupling constant, leading to a concomitant uncertainty in
the intrinsic graphene mobility. This situation is similar
\cite{Kawamura2} to the 2D
GaAs system where in fact precise measurement of the phonon-limited 2D
mobility led to the correct deformation potential coupling for
transport studies, and this could be the case for graphene also where
a quantitative comparison between our theoretical results presented in
this work with the measured temperature-dependent graphene mobility,
very recently becoming available \cite{Fuhrer2,Geim2}, could lead to
an accurate 
determination of the graphene electron-phonon deformation potential
coupling constant.

The paper is organized as follows. In section II the Boltzmann
transport theory is presented to calculate
acoustic phonon scattering limited 2D graphene conductivity. Section
III presents the results of the calculation. In  
section IV we discuss the results compared to
experimental data, and we conclude in Section V.

\section{Theory}

We use the Boltzmann transport theory
\cite{Kawamura1,Kawamura2} to calculate
acoustic phonon scattering limited 2D graphene conductivity. We
consider only the longitudinal acoustic (LA) phonons in our theory
since either the couplings to other graphene lattice phonon
modes are too weak or the energy scales of these (optical) phonon
modes are far too high for them to provide an effective scattering
channel in the temperature range ($5-500K$) of our interest.

The conductivity of graphene is given by
\begin{equation}
\sigma = e^2D(E_F)\frac{v_F^2}{2}\langle \tau \rangle,
\end{equation}
where $v_F$ is the Fermi velocity, $D(E_F) = (g_s g_v/2\pi
\hbar^2)E_F/v_F^2$ is the density of 
states of graphene at the Fermi level ($E_F$) and $\langle \tau \rangle$ is
the relaxation time averaged over energy, i.e.
\begin{equation}
\langle \tau \rangle = \frac{\int d\ve D(\ve)\tau(\ve)\left [-
  \frac{df(\ve)}{d\ve} \right ]} {\int d\ve D(\ve)\left [-
  \frac{df(\ve)}{d\ve} \right ]},
\end{equation}
where $f(\ve)$ is the Fermi distribution function, 
$f(\epsilon_k) =\{ 1+\exp[\beta(\epsilon_k-\mu)] \}^{-1}$ 
with $\beta = 1/k_BT$ and $\mu(T,n)$ as the finite temperature
chemical potential determined  
self-consistently.
The energy dependent relaxation time [$\tau(\ve_{\vk})$] is defined by
\begin{equation}
\frac{1}{\tau(\ve_{\vk})} = \sum_{\vk'}(1-\cos\theta_{\vk \vk'}) W_{\vk
  \vk'}\frac{1 - f(\varepsilon')}{1-f(\varepsilon)}
\end{equation}
where $\theta_{\vk \vk'}$ is the scattering angle between $\vk$ and
$\vk'$, $\ve = \hbar v_F |{\vk}|$, and
$W_{\vk \vk'}$ is the transition probability from the state with 
momentum $\vk$ to $\vk'$ state. In this paper we only consider the
relaxation time due to deformation potential (DP) coupled acoustic
phonon mode. The deformation potential due to 
quasi-static deformation of lattice is taken into account. Then the
transition probability has the form
\begin{equation}
W_{\vk \vk'}=\frac{2\pi}{\hbar}\sum_{\vq}|C(\vq)|^2
\Delta(\varepsilon,\varepsilon')   
\end{equation}
where $C(\vq)$ is the matrix element for scattering by acoustic phonon
and 
$\Delta(\ve,\ve')$ is given by
\begin{equation}
\Delta(\ve,\ve') = N_q \delta(\ve-\ve'+\omega_{\vq}) + (N_q + 1)
\delta(\ve-\ve'-\omega_{\vq}),
\label{delta}
\end{equation}
where $\omega_{\vq}=v_{ph} \vq$ is the acoustic phonon energy with
$v_{ph}$ being the phonon velocity  and
$N_q$ is the phonon 
occupation number
\begin{equation}
N_q = \frac{1}{\exp(\beta \omega_{\vq}) -1}.
\end{equation}
The first (second) term is Eq. (\ref{delta}) corresponds to the
absorption (emission) of an acoustic phonon of wave vector $\vq = \vk-\vk'$.
The matrix element $C(\vq)$ is independent of the phonon
occupation numbers.
The matrix element $|C(\vq)|^2$ for the deformation potential is given by
\begin{equation}
|C(\vq)|^2 = \frac{D^2\hbar q}{2A\rho_m v_{ph}}\left [ 1- \left (
    \frac{q}{2k} \right )^2 \right ],
\end{equation}
where $D$ is the deformation potential coupling constant, $\rho_m$
is the graphene mass density, and $A$ is the area of the sample.

The scattering of electrons by acoustic phonons may be considered
quasi-elastic since $\hbar \omega_{\vq} \ll E_F$, where $E_F$ is the
Fermi energy. 
There are two transport regimes, which apply to the
temperature regimes $T \ll T_{BG}$ and $T \gg T_{BG}$, depending on
whether the phonon system is degenerate (Bloch-Gr\"{u}neisen, BG) or
non-degenerate (equipartition, EP). The
characteristic temperature $T_{BG}$ is defined as $k_B T_{BG} = 2 k_F
v_{ph}$, which is given, in graphene, by $T_{BG} = 2 v_{ph}k_F/k_B \approx 54
\sqrt{n}$ K with density 
measured in unit of $n=10^{12}cm^{-2}$.  
First we consider $\hbar \omega_{\vq} \ll k_B T$. In this case we have
$N_q \sim k_B T/\hbar \omega_q$, and $\Delta(\ve,\ve') = (2k_BT/\hbar
\omega_{\vq})\delta(\ve-\ve')$. Then the relaxation time is calculated
to be
\begin{equation}
\frac{1}{\tau(\ve_{\vk})} =
\frac{1}{\hbar^3}\frac{\ve_{\vk}}{4v_F^2}\frac{D^2}{\rho_m v_{ph}^2}
k_BT.
\end{equation}
Thus, in the non-degenerate EP regime ($\hbar \omega_{\vq} \ll k_B T$) the
scattering rate [$1/\tau(\ve_{\vk})$] depends linearly on the
temperature. Since at low 
temperatures ($T_{BG} \ll T \ll E_F/k_B$) $\langle \tau \rangle
\approx \tau(E_F)$ 
the calculated conductivity is independent of Fermi energy or
equivalently the electron density. Therefore the electronic
mobility in graphene is inversely proportional to the carrier density,
i.e. $\mu \propto 1/n$. The EP regime has recently been considered in
the literature \cite{Geim2,Stauber}. We note that
the similar linear temperature dependence of the scattering time has been
reported for nanotubes \cite{Kane} and graphites \cite{phonon}.

To calculate the relaxation times in the BG regime where
$\hbar \omega_{\vq} \sim k_BT$ we have to keep the full form as in
Eq. (\ref{delta}). 
Since the acoustic-phonon energy is comparable to $k_BT$ the
temperature dependence of the relaxation time via the statistical
occupation factors in Eq. (\ref{delta}) becomes more complicated.
In BG regime the scattering rate is strongly reduced by the
occupation factors because for phonon
absorption the phonon population 
decreases exponentially and also phonon emission is prohibited by a sharp
Fermi distribution. To calculate the low temperature behavior of the
resistivity we can rewrite the averaged inverse scattering time over
energy as
\begin{equation}
\frac{1}{\langle \tau \rangle} =  \frac{1}{2\pi \hbar}
\frac{2E_F}{(\hbar v_F)^2} \int dq(1-\cos \theta) |C(q)|^2 G(\omega_q),
\end{equation}
where $q = 2k_F \sin(\theta/2)$ and $G(\omega)$ is given by \cite{Price} 
\begin{eqnarray}
G(\omega)& = &\frac{1}{k_BT} \int d\ve f(\ve)  \{ N_q
  [1-f(\ve+\omega)] \nonumber \\
& & \hspace{2cm} + (N_q+1)[1-f(\ve-\omega)] \} \nonumber \\
& = &\frac{2\omega}{k_BT}N_q(N_q+1).
\end{eqnarray}
Then we have in low temperature limits $T \ll T_{BG}$
\begin{equation}
\frac{1}{\langle \tau \rangle} \approx  \frac{1}{\pi} \frac{1}{
  E_F} \frac{D^2}{2\rho_mv_{ph}}\frac{4! \zeta(4)}{(\hbar
  v_{ph})^4}(k_BT)^4.
\label{tau_t}
\end{equation}
Thus, we find that the temperature dependent resistivity in BG regime
becomes $\rho \sim T^{4}$ without screening effects. 
If we include screening effects by the carriers themselves \cite{Hwang2}
the low-temperature resistivity
goes as $\rho \sim T^{6}$.
The screening effects on the bare scattering rates can be introduced
by dividing the matrix elements $C(\vq)$ by the dielectric function of
graphene. But the matrix elements in graphene arise from the change in
the overlap between orbitals placed on different atoms and not from a
Coulomb potential. Thus, we neglect screening effects in the
calculation, and only consider unscreened deformation potential
coupling. Even though the resistivity in EP regime is density independent,
Eq. (\ref{tau_t}) indicates that the calculated resistivity in BG regime
is inversely proportional to the density, i.e. $\rho_{BG} \sim 1/n$,
or equivalently the mobility in BG regime is density independent.

\begin{figure}
\epsfysize=2.2in
\epsffile{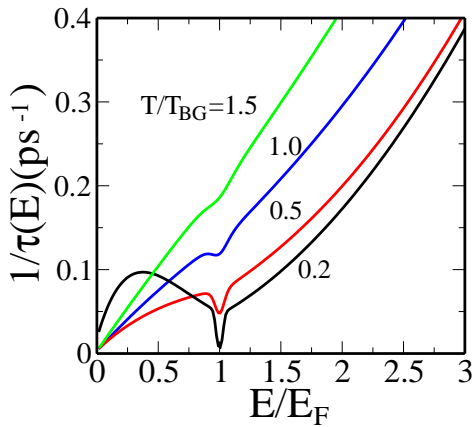}
\caption{ (Color online)
Calculated inverse relaxation times as a
function of energy for different temperatures $T/T_{BG} = 0.2$, 0.5,
1.0, and 1.5 for an electron density $n=10^{12}$ cm$^{-2}$ with
$T_{BG}=54K$. The deformation potential coupling constant $D=19$ eV
and the phonon velocity $v_{ph} = 2 \times 10^6$ cm/s are used in this
calculation. 
}
\end{figure}

\section{Results} 

In this calculation we use the
following parameters: graphene mass density $\rho_m=7.6\times 10^{-8}
\; g/cm^2$, acoustic phonon velocity $v_{ph} = 
2\times 10^6 cm/s$, and deformation potential $D = 19 eV$. 
Even though the phonon velocity $v_{ph}$ is well defined experimentally the
value of the deformation potential coupling constant is not
established \cite{Deformation,phonon}. In general, the constant $D$
could be obtained on the basis of the fact that the shift of energy
dispersion from its equilibrium state reaches the order of the atomic
energy, i.e. $D \sim e^2/a$ with $a$ being the lattice constant, which
is of the order of 10 eV in graphene. 
We note that we have used ref. \onlinecite{Deformation} for obtaining the
phonon parameters, 
but different values of $D$, differing by factors of three (i.e. $D
\approx 10-30$ eV), are quoted \cite{Deformation,phonon,Stauber} in
the literature. Since $\tau^{-1} 
\propto D^2$, the resulting graphene resistivity could differ by an
order of magnitude depending on the precise value of $D$.

In Fig. 1 we show the calculated inverse relaxation times for
deformation potential scattering by acoustic phonon as a
function of energy for different temperatures $T/T_{BG} = 0.2$, 0.5,
1.0, and 1.5 for an electron density $n=10^{12}$ cm$^{-2}$ with
$T_{BG}=54K$. The inverse relaxation time in BG regime ($T < T_{BG}$)
shows a characteristic dip (suppression of scattering rate)
in a narrow region around Fermi energy $E_F$ due to 
the statistical occupation factors. Above Bloch-Gr\"{u}neisen
temperature ($T > T_{BG}$) the dip structure disappear and the
scattering rate becomes close to the scattering rate of
equipartition regime.

\begin{figure}
\epsfysize=2.2in
\epsffile{fig2a.eps}
\epsfysize=2.2in
\epsffile{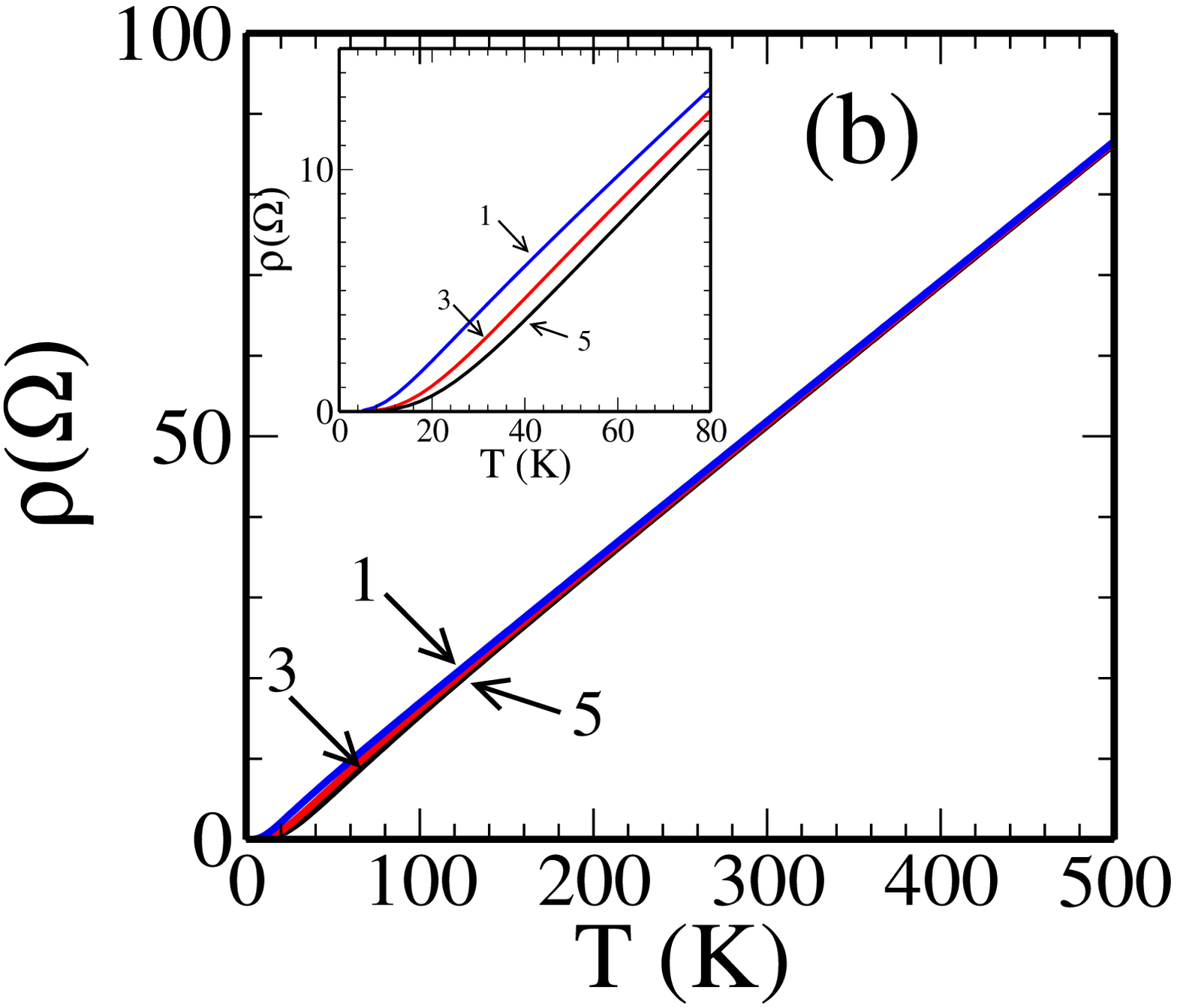}
\caption{(Color online)
(a) Calculated graphene resistivity as a function of temperature
  for several densities $n=1,$ 3, 5$\times 10^{12}$ cm$^{-2}$. We use
  the deformation potential $D=19$ eV. Note
  that in BG regime ($T<T_{BG}$)  $\rho \sim 
T^4$ and in  EP regime ($T >100K$) $\rho \sim T$.
(b) The same as Fig. 2(a) in linear scale. Inset shows the resistivity in low
temperature limits ($T<80K$).
}
\end{figure}

In Fig. 2(a) we show our calculated graphene resistivity,
$\rho \equiv \sigma^{-1}$, as a function of temperature on a log-log
plots, clearly demonstrating the two different regimes: BG $\rho \sim
T^4$ behavior for $T<T_{BG} \sim 20-100K$ and the EP $\rho \sim T$
behavior for $T >100K$. We note that $T_{BG}$ ($\propto \sqrt{n}$)
depends weakly on density, and the true BG behavior is likely to show
up at relatively low temperatures. 
In Fig. 2(b), we show $\rho(T)$ for several densities on a linear plot,
emphasizing the strong linear in $T$ dependence of the acoustic
phonon-limited graphene resistivity ranging from 100K to 500K. This
rather large temperature range of $\rho(T)\sim T$ behavior of acoustic
phonon scattering limited resistivity is quite generic to 2D
semiconductor structures, and our finding for graphene here is
qualitatively similar to what was earlier found to be the ease for 2D
GaAs structures \cite{Kawamura1,Kawamura2}. The crucial difference
between graphene and 2D GaAs 
is that in the latter system polar optical phonon scattering becomes
exponentially more important for $T \agt 100 K$ and dominates at room
temperatures whereas in graphene we predict a linear 2D resistivity
upto very high temperatures ($\sim 1000K$) since the relevant optical
phonon have very high energy ($\sim 2000K$) and are simply irrelevant
for carrier transport.

\begin{figure}
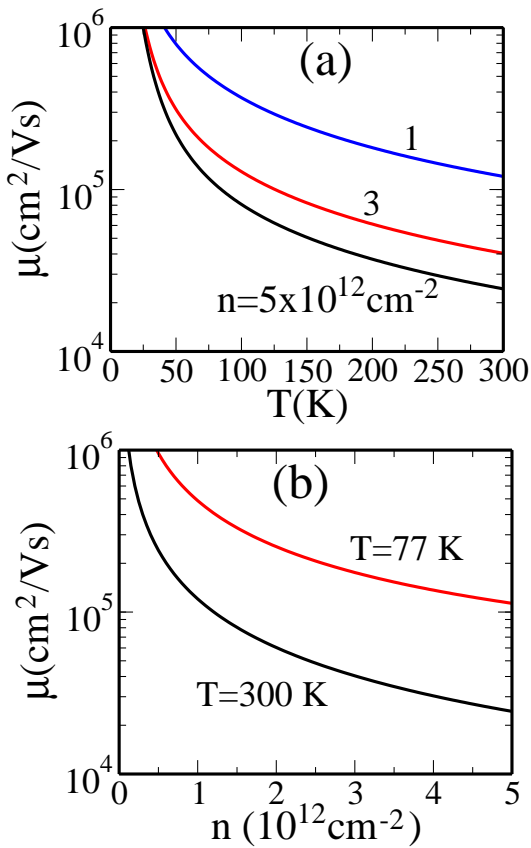

\epsfysize=2.2in
\epsffile{fig3a.eps}
\epsfysize=2.2in
\epsffile{fig3b.eps}
\caption{ (Color online)
Calculated mobility limited by the acoustic phonon with the
deformation potential coupling constant $D=19$ eV (a) as a function of
temperature for different densities $n=1$, 3, 5$\times 10^{12}$
cm$^{-2}$ and (b) as a function of density for different temperatures
T=77K and T=300K.
}
\label{mu_ph}
\end{figure}

In Fig. 3, we show our calculated intrinsic graphene mobility, $\mu
\equiv (en\rho)^{-1}$, as functions of temperature and carrier
density. Within our model, the unscreened acoustic phonon scattering
limited graphene mobility is inversely proportion to $T$ and $n$
individually for $T \agt 100K$. Assuming $D=19 eV$, as used in Fig. 3,
$\mu$ could reach values as high as $10^5$ cm$^2/Vs$ for lower carrier
densities ($n \alt 10^{12}$ cm$^{-2}$). For larger (smaller) values of
$D$, $\mu$ would be smaller (larger) by a factor of $D^2$. It may be
important to emphasize here that we know of no other system where the
intrinsic room-temperature carrier mobility could reach a value as
high as $10^5$ cm$^2/Vs$. 

\section {Discussion}

The three key theoretical findings on phonon-limited
graphene mobility of this work are : (1) $\rho \sim T$ for $T \agt
100K$; (2) $\rho \propto T^4$ ($T^6$) for $T \alt 50K$) for
unscreened (screened) deformation potential coupling; (3) $\mu \agt
3.7\times 10^7/D^2 \tilde{n}$ cm$^2/Vs$ at room temperature ($T=300K$) where $D$
is measured in eV and $\tilde{n}$ is carrier density $n$ measured in
units of $10^{12}$ cm$^{-2}$. These theoretical predictions being
rather precise, the question naturally arises about the experimental
status and the verification of our theory. Very recently, experimental
graphene transport data at room temperature (or even above
\cite{Fuhrer2}) have 
started becoming available \cite{Fuhrer2,Geim2}. The only aspect of
our theory that can be 
directly compared with the existing experiment is the $\rho \propto T$
behavior at high temperatures ($ > 200K$), and this is indeed
consistent with the recent data from two different groups
\cite{Fuhrer2,Geim2}. Geim and 
collaborators have recently concluded \cite{Geim2} that the room
temperature 
graphene intrinsic mobility could be as high as $10^5$ cm$^2/Vs$,
which is also consistent with our theory. But, as emphasized by us,
the actual mobility value varies inversely as $D^2$, and therefore a
precise knowledge of the deformation potential coupling is required
for an accurate estimate of the intrinsic mobility.

A more detailed comparison between our theory and the experimental
results on $\rho(T)$ shows some qualitative difference which are not
understood at this point. For example, the experimental crossover
\cite{Fuhrer2,Geim2} to
the high-temperature linear ($\rho \sim T$) behavior in the intrinsic
resistivity appears to be closer to a $T^2$ behavior rather than the
$T^4$ BG behavior we predict. More disturbingly, the experimental
crossover from the high-temperature linear behavior to the
low-temperature high power law behavior appears to be occurring at a
much higher temperature ($100-200K$) than the theoretical prediction
($20-50K$). At this stage we have no explanation for the
lower-temperature disagreement between experiment and theory, but
below we discuss several possibilities.

\begin{figure}
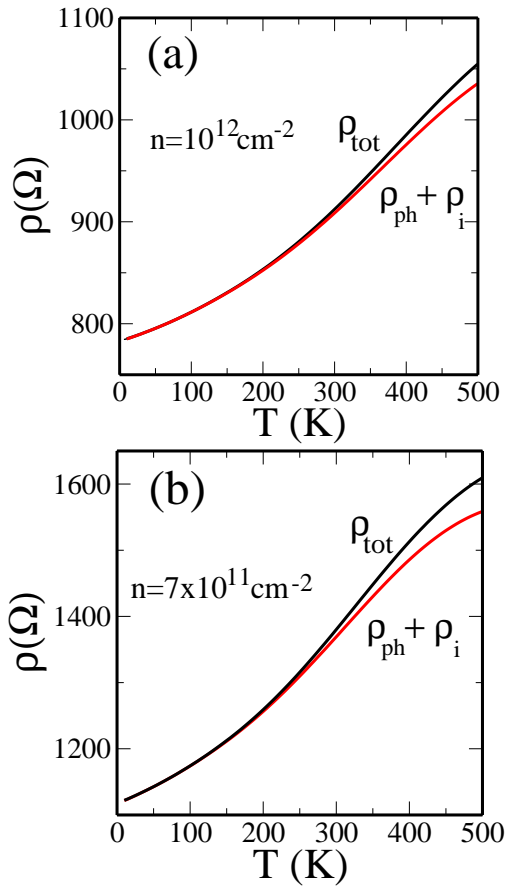

\epsfysize=2.3in
\epsffile{fig4a.eps}
\epsfysize=2.3in
\epsffile{fig4b.eps}
\caption{ (Color online)
The deviations from Mattiessen's rule for two different densities, (a)
$n=10^{12}$ cm$^{-2}$ and (b) $n=7\times 10^{11}$ cm$^{-2}$. Here
$\rho_i$ ($\rho_{ph}$) represents the resistivity due to impurity
(phonon) scattering. 
}
\label{mat}
\end{figure}

The experimentally measured resistivity in the current graphene
samples is completely dominated by extrinsic scattering (and {\it not}
by phonon scattering) even at room temperatures since the
low-temperature ($\alt 4K$) mobility is typically $5,000-15,000$
cm$^2/Vs$, and the intrinsic room-temperature phonon contribution, as
obtained theoretically by us or inferred \cite{Fuhrer2,Geim2} from recent
temperature-dependent experiments, is $10-20$ times larger ($\sim
100,000$cm$^2/Vs$). This means that any experimental extraction of
the pure phonon contribution to graphene resistivity involves
subtraction of two large resistances (i.e. the measured total
resistance and the extrapolated $T=0$ extrinsic temperature-independent
resistance arising from impurity and defect scattering) of the order
of $k\Omega$ each to get a phonon contribution roughly of the order of
100 $\Omega$. Apart from the inherent danger of large unknown errors
involved in the subtraction of two large numbers to obtain a much
smaller number associated with phonon scattering contribution to
graphene mobility, there is the additional assumption of the
Matthiessen's rule, i.e. $\rho_{tot} = \rho_{ph}+\rho_{i}$ where
$\rho_{tot}$ is the total resistivity contributed by impurities and
defects ($\rho_{i}$) and phonons ($\rho_{ph}$), which is simply not
valid. In particular, the impurity contribution to resistivity also
has a temperature dependence arising from Fermi statistics and
screening which, although weak, cannot be neglected in extracting the
phonon contribution (particularly since the total phonon contribution
itself is much smaller than the total extrinsic contribution). In
particular, the temperature dependent part of the charged impurity
scattering contribution to graphene resistivity could be positive or
negative \cite{Hwang3} depending on whether screening or degeneracy effects
dominate, and therefore the phonon contribution, as determined by a
simple subtraction, could have large errors, particularly in the low
($T<1000K$) temperature regime. Indeed, a recent measurement
\cite{Kim_T} of
$\rho(T)$ in the $0-100K$ regime finds small temperature dependent
contributions to graphene resistivity which could be either positive
or negative depending on the sample mobility and which, in all
likelihood, arises from extrinsic impurity scattering.

In Fig. 4 we show the failure of the Matthiessen's rule in graphene
(particularly at higher/lower temperatures/densities) by calculating
the total graphene resistivity arising from screened charged impurity
scattering ($\rho_i$) and phonon scattering ($\rho_{ph}$) --- it is
clear that $\rho_{tot} \neq \rho_{ph} + \rho_i$ for lower/higher
densities/temperatures.

If the experimentally extracted \cite{Fuhrer2,Geim2} phonon
contribution to the graphene 
resistivity turns out to be accurate in spite of the rather questionable
subtraction procedure discussed above, then the disagreement between our
intermediate temperature (50---100K) theoretical results and the experimental
data would indicate the presence of some additional phonon modes which must
be participating in the scattering process.  We can, in fact, get reasonable
agreement between our theory and the experimental data by arbitrarily
shifting the BG temperature $T_{BG}$ to a higher temperature around
200K.  This shift could indicate a typical phonon scale which causes
additional scattering other than the LA phonons coupled to the carriers
through the deformation potential coupling considered in our work.  At this
stage we cannot speculate on what these additional modes could be.  One
possibility is that these are the zone-edge out of plane ZA phonon 
modes with vibrations transverse to the graphene plane \cite{Fuhrer2}.  
(In the Appendix we provide the calculated carrier resistivity in the
presence of an additional phonon mode with a soft gap.)
Another possibility
considered in ref. \onlinecite{Geim2} is that these are the thermal
fluctuations 
("ripplons") of the mechanical ripples invariably present in graphene
samples \cite{Adam}.  
Of course such additional ``phonon'' scattering channels will lead to
additional unknown coupling parameters making the resultant theory
essentially a data fitting procedure. The advantage of our minimal
theory is that it involves only two phonon parameters: $D$ and
$v_{ph}$ associated with the 2D graphene LA phonons.
More data in higher mobility samples will
be needed to settle this question since the subtraction problem inherent in
the current technique for extracting the phonon contribution would make
analyzing this issue a difficult task.

\section{Conclusion}

We have calculated the intrinsic temperature
dependent 2D graphene transport behavior upto 500K by considering
temperature and density dependent scattering of carriers by acoustic
phonons. We have provided a critical discussion of our results in
light of the recent experiments \cite{Fuhrer2,Geim2}.
The lack of precise quantitative knowledge about graphene deformation
potential coupling makes a quantitative comparison with the
experimental data problematic.

We thank Michael Fuhrer (ref. [\onlinecite{Fuhrer2}]) and Andre Geim
(ref. [\onlinecite{Geim2}]) for sharing with 
us their unpublished data, and Andre Geim for a careful reading of our
manuscript.
This work was supported by U.S. ONR, LPS-NSA, and SWAN-NSF-NRI.

\begin{figure}
\epsfysize=2.5in
\epsffile{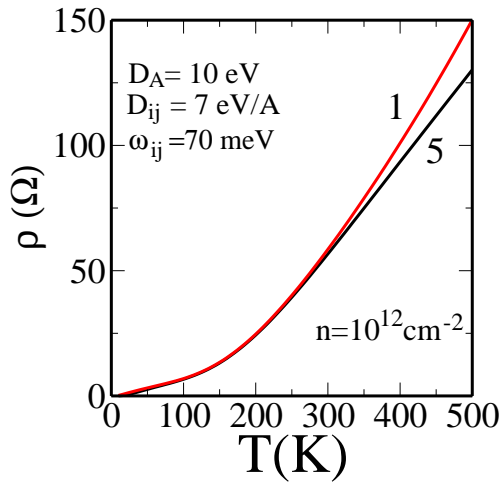}
\caption{ (Color online)
Calculated resistivity with both the acoustic phonon scattering and
the inter-valley phonon scattering. See appexdix for details.
}
\end{figure}

\section*{APPENDIX}

In this appendix we calculate the phonon scattering limited carrier
mobility including effects of two phonon branches: the regular LA
phonon (as considered in the main part of this paper) and an
additional ``intervalley'' phonon branch with a soft gap ($\sim 70$
meV) representing the inter-valley phonon, the ZA phonon mode at the K
point. The theoretical motivation is to demonstrate that the
combination of the LA phonon and an optical phonon (i.e. the
inter-valley ZA mode) with a softy gap could indeed lead to qualitative
(or even quantitative, if the phonon parameters have appropriate
values) agreement between theory and experiment.

In this context we consider the long wavelength phonon scattering.
As temperature increases phonons of large wave vectors are involved in
the scattering in multi-valley structures. Thus the inter-valley phonon
scattering becomes significant at high temperatures \cite{Fuhrer2}. 
In graphene there are two minima of the conduction band at
K and K' points in Brillouin zone. 
The scattering between K and K' points requires the participation of
inter-valley 
phonons, whose wave vectors are close to {\bf q}$_{ij} = {\bf k}_K-
{\bf k}_{K'}$ and the frequencies of 
these phonons are close to $\omega_{ij} = \omega(q_{ij})$.
The relaxation time for inter-valley phonon scattering may be
considered by assuming constant inter-valley phonon energies $\hbar
\omega_{ij}$. Then the matrix element for the inter-valley
phonon scattering becomes 
$|C(\vq)|^2 = {\hbar D_{ij}^2}/{2A\rho_m \omega_{ij}}$
where $D_{ij}$ is the deformation potential coupling constant for
inter-valley phonons in unit of eV/\AA.
Since $\omega_{ij} \ll E_F$ in graphene the scattering of electrons
from inter-valley phonons is considered quasielastically.

In Fig. 5 we show the calculated resistivity with both the acoustic phonon
scattering and the inter-valley phonon scattering as a function of
temperature for two different densities.
The following parameters are used in this calculation: deformation
potential coupling constant $D=10 eV$, acoustic phonon velocity 
$v_{ph} = 2\times 10^{6}$ cm/s, $D_{ij} = 7$ eV/\AA, and inter-valley
phonon energy $\hbar \omega_{ij} =
70$ meV which corresponds to the lowest phonon energy (ZA) at the K point. 
The calculated resistivities have very weak density dependence, which
comes from the energy averaging. Below
200K the acoustic phonon scattering dominates (linear in temperature),
but above 200 K both phonon scatterings contribute in the
transport. Note that the temperature dependence of the 
resistivity is linear in both regimes, but has different slopes. The
high power law behavior ($\rho \sim T^4$) only applies at very low
temperatures ($T < 50$ K).   
Basically, there is a sharp turn-on in phonon scattering in the
150-250K range as the inter-valley ZA phonon scattering becomes
effective.

Whether the results shown in Fig. 5, including the effects of
inter-valley phonon scattering are physically meaningful or not will
depend on the direct observation of these ZA phonon modes via Raman
scattering experiments. At this stage, all we have established is that
inclusion of this additional soft-gap phonon mode gives impressive
agreement between theory and experiment \cite{Fuhrer2,Geim2}. More
work is needed to validate the model of combined LA and ZA phonon
scattering contributing to the temperature dependent graphene
resistivity.

\end{document}